\documentclass[12pt, preprint]{aastex}
\usepackage{natbib,graphicx,amsmath,amsthm,ulem, subfigure, enumitem, color}
\definecolor{hypercolor}{RGB}{0,0,127}
\usepackage[%
  citecolor=hypercolor,%
  linkcolor=hypercolor,%
  urlcolor=hypercolor,%
  backref=false,%
  pagebackref=false%
]{hyperref}%

\newcommand{\documentname}{\textsl{white paper}}

\newcommand{\observatory}[1]{\textsl{#1}}
\newcommand{\kepler}{\observatory{Kepler}}
\newcommand{\Kepler}{\kepler}
\newcounter{hoggitem}
\newcounter{address}

\shorttitle{Save Kepler 2}
\shortauthors{Montet et al.}

\begin{document}\sloppy\sloppypar\thispagestyle{empty}

\title{Maximizing \Kepler\ science return per telemetered pixel: \\
  Searching the habitable zones of the brightest stars\altaffilmark{\ref{kcall}}}

\author{%
  Benjamin~T.~Montet\altaffilmark{\ref{Caltech},\ref{email}},
  Ruth~Angus\altaffilmark{\ref{Oxford}},
  Tom~Barclay\altaffilmark{\ref{Ames}},
  Rebekah~Dawson\altaffilmark{\ref{CfA}},
  Rob~Fergus\altaffilmark{\ref{Courant}},
  Dan~Foreman-Mackey\altaffilmark{\ref{CCPP}},
  Stefan Harmeling\altaffilmark{\ref{MPIIS}},
  Michael~Hirsch\altaffilmark{\ref{UCL},\ref{MPIIS}},
  David~W.~Hogg\altaffilmark{\ref{CCPP},\ref{MPIA}},
  Dustin~Lang\altaffilmark{\ref{CMU}},
  David~Schiminovich\altaffilmark{\ref{Columbia}} \&
  Bernhard~Sch\"olkopf\altaffilmark{\ref{MPIIS}}%
}

\setcounter{address}{1}
\altaffiltext{\theaddress}{\stepcounter{address}\label{kcall}%
  A \documentname\ submitted in response to the \Kepler\ Project Office
  \textit{Call for White Papers: Soliciting Community Input for
    Alternate Science Investigations for the Kepler Spacecraft}
  released 2013 August 02
  (\url{http://keplergo.arc.nasa.gov/docs/Kepler-2wheels-call-1.pdf})}
\altaffiltext{\theaddress}{\stepcounter{address}\label{Caltech}%
  Department of Astronomy, California Institute of Technology}
\altaffiltext{\theaddress}{\stepcounter{address}\label{email}%
  To whom correspondence should be addressed; \texttt{<btm@astro.caltech.edu>}.}
\altaffiltext{\theaddress}{\stepcounter{address}\label{Oxford}%
  Department of Physics, Oxford University}
\altaffiltext{\theaddress}{\stepcounter{address}\label{Ames}%
  NASA Ames Research Center}
\altaffiltext{\theaddress}{\stepcounter{address}\label{CfA}%
  Harvard--Smithsonian Center for Astrophysics}
\altaffiltext{\theaddress}{\stepcounter{address}\label{Courant}%
  Courant Institute of Mathematical Sciences, New York University}
\altaffiltext{\theaddress}{\stepcounter{address}\label{CCPP}%
  Center for Cosmology and Particle Physics, Department of Physics, New York University}
\altaffiltext{\theaddress}{\stepcounter{address}\label{MPIIS}%
  Max-Planck-Institut f\"ur Intelligente Systeme, T\"ubingen}
\altaffiltext{\theaddress}{\stepcounter{address}\label{UCL}%
  Department of Physics and Astronomy, University College London}
\altaffiltext{\theaddress}{\stepcounter{address}\label{MPIA}%
  Max-Planck-Institut f\"ur Astronomie, Heidelberg, Germany}
\altaffiltext{\theaddress}{\stepcounter{address}\label{CMU}%
  McWilliams Center for Cosmology, Carnegie Mellon University}
\altaffiltext{\theaddress}{\stepcounter{address}\label{Columbia}%
  Department of Astrophysics, Columbia University}

% comment
\newpage
~
\clearpage

% Astro-Ph Abstract:
% In today's mailing, Hogg et al. propose image modeling techniques to maintain 10-ppm-level precision photometry in Kepler data with only two working reaction wheels. While these results are relevant to many scientific goals for the repurposed mission, all modeling efforts so far have used a toy model of the Kepler telescope. Because the two-wheel performance of Kepler remains to be determined, we advocate for the consideration of an alternate strategy for a >1 year program that maximizes the science return from the ``low-torque'' fields across the ecliptic plane. Assuming we can reach the precision of the original Kepler mission, we expect to detect 800 new planet candidates in the first year of such a mission. Our proposed strategy has benefits for transit timing variation and transit duration variation studies, especially when considered in concert with the future TESS mission. We also expect to help address the first key science goal of Kepler: the frequency of planets in the habitable zone as a function of spectral type.

%A white paper submitted in response to the "Kepler Project Office Call for White Papers: Soliciting Community Input for Alternate Science Investigations for the Kepler Spacecraft"; 14 pages in length (that is, a modest 4 pages over the white-paper page limit).
\section{Executive Summary}

\paragraph{Primary Eecommendation:}
In Paper I (Hogg et al.) we propose image modeling techniques to 
  maintain 10-ppm-level precision photometry in \Kepler\ data with only two working 
  reaction wheels. 
While these results are relevant to many scientific goals
  for the repurposed mission, all modeling efforts so far have used a toy 
  model of the \Kepler\ telescope. 
Because the two-wheel performance of \Kepler\ remains to be determined, 
 \textbf{we 
 advocate for the consideration of an alternate strategy for a $>1$ year program
 that maximizes the science return from the ``low-torque'' fields across the
 ecliptic plane.}
There are considerable benefits of such a strategy which make this design a 
  viable approach for \Kepler\ in any scenario, but especially one in which, 
  for any reason, the field analyzed in the primary mission cannot be used 
  moving forward.
\begin{itemize}
\item
By an analysis of planetary candidates previously detected by \Kepler, 
 if we can achieve photometry equal to that of the primary
 mission we conservatively expect to detect 800 new planetary candidates 
 (after a thorough vetting for false positives). Even if our photometry is 
 degraded by a factor of two from the primary mission, our conservative
 estimate decreases only to 400 new planetary candidates.
\item
The first scientific goal\footnote{\url{http://kepler.nasa.gov/science/about/scientificgoals/}} of \Kepler\ is to \textbf{determine the frequency 
 of terrestrial and larger planets in or near the habitable zone of a wide 
 variety of spectral types of stars}. 
Not only does this recommendation not detract from this strategy, it may 
 provide the best chance to answer this question moving forward.
\item
In Hogg et al. we argue for a shorter target list. 
By culling the target list such that we are limiting ourselves to bright stars,
 the \Kepler\ long cadence integrations of 29.4 minutes can also be shortened,
 which has the benefit of 
 allowing for more sensitive observations of transit timing and transit
 duration variations.
These stars will be later observed by the Transiting Exoplanet Survey Satellite
 \citep[TESS][]{Brown08}, but only for 30 days; there will 
 not be enough transits observed in this mission to detect variations due to 
 planet-planet interactions.
Therefore, by observing bright stars that can be followed up by TESS, this 
 will enable dynamical studies that could not be undertaken with either 
 telescope alone.
\item
Similarly, shorter integrations may allow for asteroseismic studies of 
 more stars. 
Most asteroseismic targets have been giant stars, for which 
 the period of oscillations are long enough to be observed in 
 long cadence data \citep{Chaplin13}.
By decreasing the integration time, solar-type oscillations can be observed 
 on subgiants and other stars nearer the main sequence.
On these stars, asteroseismic signals are smaller, but by focusing on 
 bright stars we expect the signals to be observable on stars nearer the 
 main sequence.
\item
Bright stars are also advantageous in that they allow radial velocity follow-up
 observations to be carried out. 
By selecting brighter stars, we assure interesting objects can be characterized 
 from the ground \textit{before} the launch of JWST.
 
\end{itemize}
This white paper is organized as follows. 
In \textsection\ref{Targets}, we outline our target selection and observing 
 strategies. 
In \textsection\ref{Haul}, we project the expected number of planets detected 
 by this strategy. 
In \textsection\ref{HZ}, we explain how this strategy will enable completion of 
 the primary scientific goal of the \Kepler\ mission.
Section \ref{Dynamics} outlines the advantageous of such a strategy for
 transit timing and transit duration variation studies, especially in concert
 with TESS. 
Section \ref{AS} discusses other advantages of such an observing strategy. 
We summarize and conclude in \textsection\ref{SC}. 

\section{Target Selection}
\label{Targets}
Simulations suggest pointing errors from a two-wheel \Kepler\ will be 
 minimized by pointing in the ecliptic plane, where the torque exerted about 
 the spacecraft's X and Y axes by solar pressure is approximately zero. 
Such areas are expected to be stable for approximately six months, so that the 
 expected drift is negligible over a 30 minute or shorter integration.
We propose to observe fields in the ecliptic plane for this reason. 
To avoid the Earth crossing the field of view, it is recommended the telescope
 point only in the direction opposite motion, meaning each field is stable 
 \emph{and observable} for only three months.
\textbf{Fortunately, the ecliptic plane is ideal for the \Kepler\ telescope}.
Recall the \Kepler\ field is situated 8-18 degrees out of the galactic 
 plane\citep{Gilliland11}.
This location provides a sufficient quantity of 
 $K_p < 16$ stars to observe without a prohibitive amount of crowding 
 in the telescope's 4 arcsecond pixels. 
The ecliptic plane and galactic plane cross twice, at right ascension of 
 approximately 7 and 19. 
It is therefore possible to choose four fields, each separated by
 approximately 6 hours in RA, that are both on the ecliptic plane and
 approximately the same distance from the galactic plane as the \Kepler\
 field. 
These are located near RA of 3, 9, 15, and 21. 

In this paper, we do not attempt to choose specific fields, but instead
 simply show that there are sufficient observable stars such that our proposed
 observations are feasible.
In part, this is because (as Hogg et al. discuss) our image modeling techniques
 may be more successful when the telescope drift rate is larger. 
With degraded pointing and additional drift, the diversity of stars that touch
 different combinations of pixels is increased. 
More simulations are required to determine the optimal positioning of the 
 telescope with respect to the ecliptic so that the drift rate is large enough 
 to maximize our abilities to model pixel sensitivities but small enough 
 that the stars stay within their apertures between telescope repointings.
Fortunately, it is quite simple to select four fields that are both near the 
 ecliptic and well separated such that each can be observed for approximately
 three months.

We recommend shorter integration times than the long cadence observations, 
 as we discuss more fully below. 
As a result, fewer targets are available for observations than have been 
 previously observed. 
We aim for approximately \emph{thirty thousand targets per field}, with 
 preference given to brighter stars. 
Here, ``brighter'' is intentionally left as an ambiguous term. If our goal is to detect Earth-sized planets, then M dwarfs should be allowed
 to be fainter than G dwarfs because of their deeper transit signal, for
 example. 

If our image modeling techniques are successful, then a field similar to the 
 original \Kepler\ field with respect to the galactic plane will contain 
 a sufficient number of viable solar-type and smaller targets. 
To verify each field will contain enough stars, we simulate the population 
 of the nearby galaxy with TRILEGAL \citep{Girardi05}. 

We first consider the current \Kepler\ field. 
TRILEGAL is limited to fields of 10 square degrees. 
To combat this, we simulate three regions of 8 square degrees across the 
 \Kepler\ field orthogonal to the galactic plane. 
We then estimate the extinction in the galaxy by calibrating the extinction 
 value at infinity using the values found in the extinction map created by 
 \cite{Schlegel98}. 
We return the \Kepler\ magnitude of each star in our field and scale our
 total star count to what would be expected over 100 square degrees.

\begin{figure}
\includegraphics[width=\textwidth]{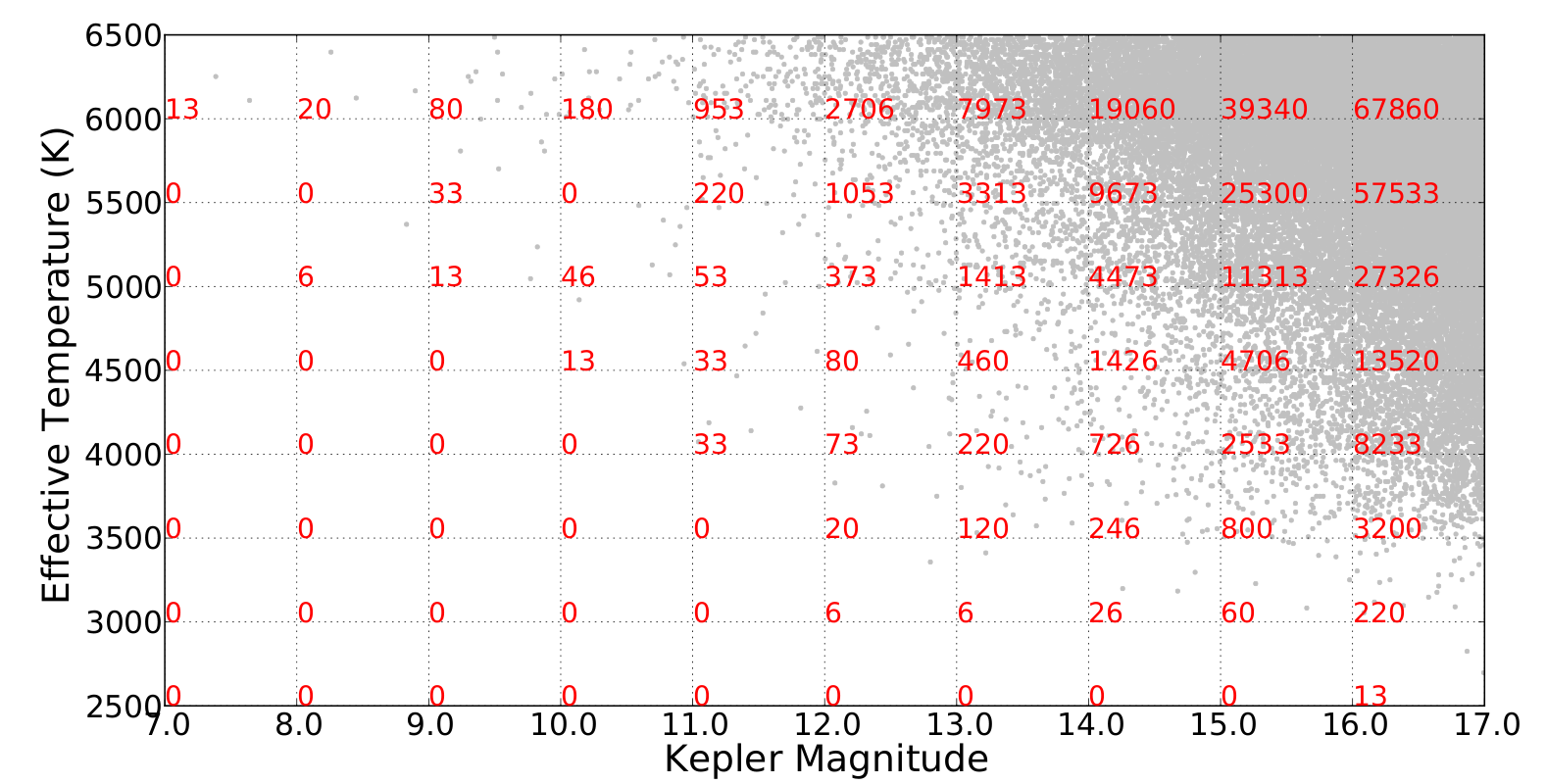}
\caption{Estimated number of visible targets in the \Kepler\ field as a 
 function of apparent magnitude in the \Kepler\ bandpass and stellar 
 effective temperature, simulated using TRILEGAL. 
As might be expected, a plurality of targets are found in the top right corner,  where luminous, distant stars exist. 
There are a sufficient number of bright targets equal to or smaller than the size
 of the Sun, around which small planets in the habitable zone could be 
 detected.
}
\label{Fig1}
\end{figure}

Fig. \ref{Fig1} shows the distribution of stars as a function of effective 
 temperature and apparent magnitude. 
Ideally, stars in the lower left region of the figure, relatively speaking,
  would be chosen. 
The number of simulated stars found is comparable with the true number of 
 observable stars in the \Kepler\ field.

\begin{figure}
\includegraphics[width=\textwidth]{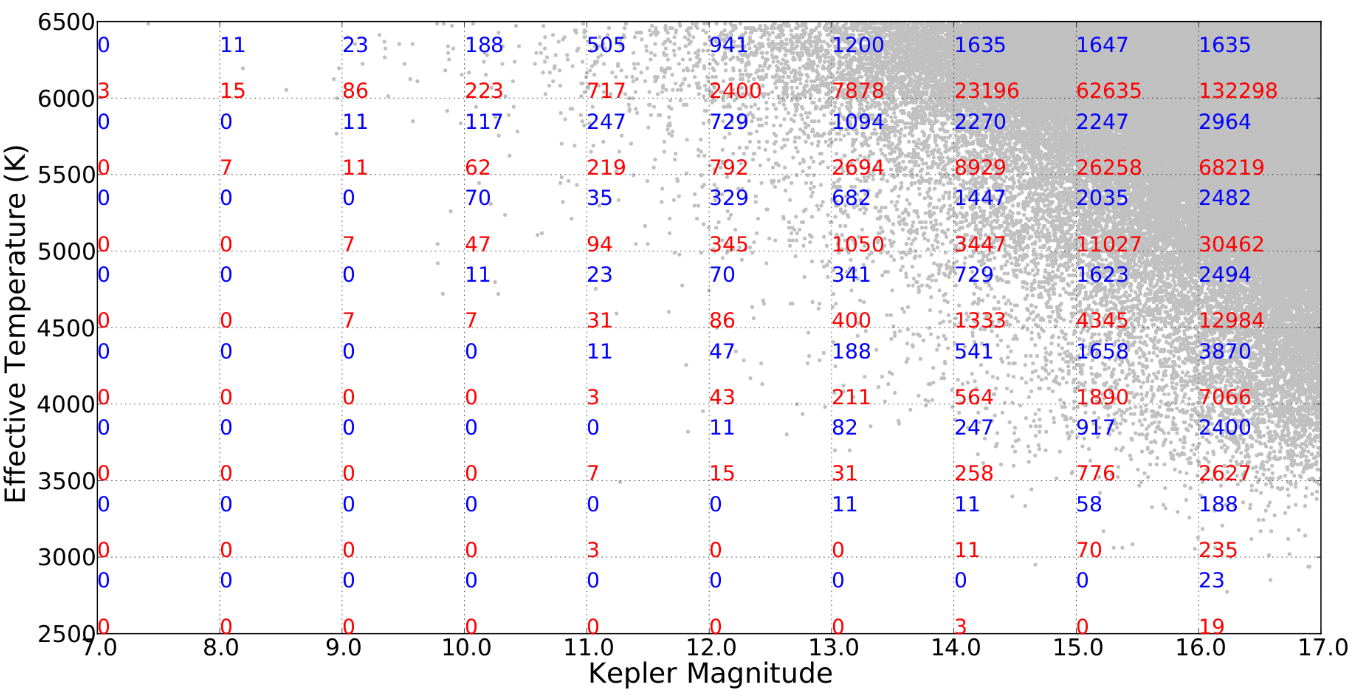}
\caption{Same as Fig. \ref{Fig1}, but for \Kepler\ field-sized patches of sky
 well out of the galactic plane (blue) and in the galactic plane, opposite the 
 galactic center (red). 
Far away from the galactic plane, there are not a sufficient number of stars 
 in one 100 square degree field of view. 
In the plane, crowding is expected to be a significant concern, inflating the 
 false positive occurrence rate. 
Therefore, we recommend selecting four fields 10-20 degrees from the galactic 
 plane, similar to the current field. 
The optimal distance from the ecliptic plane with respect to the telescope 
 drift rate has not been precisely quantified, but it is expected that four 
 appropriate fields, approximately 6 hours apart in right ascension, should 
 easily be selectable.
For a survey of M dwarfs, target selection is much less of a concern.
Note all three fields have similar numbers of low-mass stars. 
As M dwarfs are significantly underluminous relative to solar mass stars, 
 bright M dwarfs must be located within tens of parsecs and are thus
 approximately isotropic over the sky.
}
\label{Fig2}
\end{figure}

There are sufficient hypothetical fields near both the ecliptic and galactic 
 planes to develop four similar fields from which 30,000 stars could each 
 be selected. 
It is of considerable importance, however, that \textbf{if new fields are 
 selected, they should be located a similar distance from the galactic plane}. 
We provide evidence in support of this claim in the form of Fig. \ref{Fig2}. 
This figure was developed in an identical manner to Fig. \ref{Fig1}, but 
 the numbers correspond to the expected yield in a region of the sky far away
 from the galactic plane (0:00:00, +0:00:00, blue) and very close to the center
 of the galactic plane, pointing away from the galactic center (6:00:00, 
 +23:30:00, red). 
Both of these locations are on the ecliptic plane.
Out of the galactic plane, the number of solar-type stars is decremented by
 an order of magnitude relative to the \Kepler\ field.
In the center of the plane, such stars are a factor of two more common than in
 the \Kepler\ field, meaning crowding
 (and background eclipsing binary false positives) would be even more a concern
 than in the original mission\footnote{This is not a problem for future
 missions focused on the study of M dwarfs: because the stars are intrinsically
 faint, most M dwarfs brighter than 16th magnitude are within 100 parsec
 and distributed approximately isotropically across the sky.}.
Therefore, if new fields are selected we \emph{strongly recommend} selecting 
 four fields near the ecliptic plane and approximately 15 degrees from the 
 galactic plane.

\section{Expected Planet Detections}
\label{Haul}
To determine the number of planets we would expect to find with our new 
 observing strategy, we consider the best source of information on transiting 
 exoplanets, the list of \Kepler\ planet candidates found on the NASA 
 Exoplanet Archive\footnote{\url{http://exoplanetarchive.ipac.caltech.edu}}
 \citep{Akeson13}.
The number of detections depends on the level of success of our image modeling
 efforts, which has not yet been precisely quantified. 
By our mission design strategy, we propose staring at each of four fields for 
 one quarter of a year, or 93 days. 
To observe three transits of a planet, we are then limited to planets with 
 periods of 30 or fewer days. 
We therefore search the Q1-Q12 data release for planets with periods shorter 
 than 30 days. 
To simulate one quarter of observations and a 7.1$\sigma$ signal 
 to noise cut, we only select objects in the Q1-Q12 catalogue with signal to 
 noise larger than $7.1\sqrt{12}$. 
We limit ourselves to objects with $K_p < 13$, of which there are 28,000 in 
 number, to simulate our recommendation of searching 30,000 bright stars.
Finally, since we plan to observe four such fields, we multiply the number of 
 ``detections'' by four. 
We find that we expect to detect \textbf{836} candidate planets in a one year
 mission. 
As we are analyzing the \Kepler\ planet \emph{candidates}, 836 is the number 
 of candidates we expect to find \emph{after} considerable vetting efforts; 
 known false positives have been excluded from this sample.

We repeat these experiments with multiple signal to noise tolerances, 
 simulating hypothetical scenarios where our best achievable photometry was 
 degraded by some factor relative to the photometry of the prime \Kepler\ 
 mission. 
The results are shown in the table below.
\begin{center}
\begin{tabular}{|l|c|c|c|c|c|}
    \hline
      %  \tabletypesize{\footnotesize}
     %   \tablecolumns{6}
     %   \tablecaption{Expected Planet Detections}
      %  \tablehead{ Degrading Factor & 1 & 2 & 3 & 4 & 5}
      %  \startdata
    Degradation Factor & 1 & 2 & 3 & 4 & 5\\
    \hline
    Number of Detections  & 836 & 428 & 284 & 216 & 176  \\
    \hline
     %   \enddata
\end{tabular}
\end{center}
Thus, even if our photometry is a factor of three worse than the original 
 \Kepler\ mission, we still expect to detect nearly 300 new planets. 

These estimates are \textit{conservative and approximate}. 
First off, they assume a flat magnitude limit of $K_p = 13$, which would bias 
 our sample towards massive stars. 
To observe a planet with a given S/N ratio around a massive star, a
 correspondingly massive planet would be required. 
By preferentially selecting smaller stars, we are more sensitive to smaller 
 planets which are far more common \citep{Morton13}. 
Therefore, a more careful target selection will likely increase our 
 number of detections.

Additionally, the Q1-Q12 planet search was carried out by the \Kepler\ team 
 using a now-dated planet search pipeline. 
Recent improvements to this pipeline are likely to improve the sensitivity of 
 the search by an as yet unquantified amount, boosting our planet detection
 numbers even more. 

Moreover, this thought experiment assumes a one year new mission. 
If the remaining two reaction wheels encounter no problems, we would expect 
 to be able to push our detections to longer periods. 
As an added bonus, we would be biased toward planets with approximately one
 year periods due to our unique observing cadence. 
The proposed cadence is both a blessing and a curse for finding planets in the 
 so-called habitable zone, as we explain in the next section.

\section{The Frequency of ``Habitable-Zone'' Planets}
\label{HZ}
Goal 1 of the \Kepler\ mission is the determine the frequency of terrestrial 
 and larger planets in or near the habitable zone of a wide variety of 
 spectral types of stars. 
For M dwarfs the habitable zone varies considerably by spectral type,
 but for the majority of M dwarfs objects in the habitable 
 zone have periods shorter than 30 days \citep{Selsis07, Kopparapu13}.
Thus, in a one-year mission we would expect to determine the frequency of 
 M dwarf habitable systems. This number is presently poorly constrained due 
 to the small number of M dwarfs observed in the primary mission; we hope
 this is given prime consideration in the future plans for the telescope.

The true frequency of habitable zone planets is less constrained for solar-type 
 stars.
If our image modeling techniques are successful, such a strategy as the 
 one proposed here will enable an estimate of this value in a multi-year 
 extended mission. 
Because we return to the same field every year, we will be biased towards 
 planets that transit once per year. 
Planets that transit exactly twice (or three, or $N$) times per year will 
 also appear in our sample as habitable zone ``impostors.'' 
Such impostors can be accounted for in a statistical sense by an analysis of 
 transit durations (which increase as a function of period), and for 
 individual highly interesting systems by future space-based follow up. 
Therefore, we do not expect such impostors to be a significant hindrance.

For each field, because of our cadence, we expect to miss a significant 
 fraction (approximately $75\%$) of the transiting habitable zone planets. 
This is an unavoidable effect from this observing strategy, but should 
 not affect our results. 
Because \Kepler\ would always be staring at one of the four proposed fields, 
 it will always be on the lookout for \emph{some} transiting, potentially 
 habitable systems; this missed systems can be accounted for statistically. 

Thus, from this observing strategy, we expect to increase both the number 
 of Earth-like planet candidates and the precision to which planet occurrence 
 rates are measured.

\section{Dynamics with \Kepler\ and TESS}
\label{Dynamics}
\subsection{The Flatness of Transiting Exoplanetary Systems}
Despite the tremendous success of \Kepler\ at finding planets, questions remain
 as to the flatness of the exoplanetary systems uncovered. 
While the vast majority of transiting systems must have inclinations of a 
 few degrees or fewer, it is unclear how flat is ``flat.'' 
Moreover, in special cases we may expect inclined companions to transiting 
 planets. 
For example, hot Jupiters may be formed by early dynamical interactions with a 
 mutually inclined perturber, initially exchanging inclination for eccentricity
 and then circularizing through tidal effects. 
If this scenario is accurate, then we would expect inclined companions to 
 slowly perturb transiting planets. 
As the inclination of a transiting planet changes, so does the chord the planet
 cuts across the face of its star. This also necessarily changes the 
 \emph{duration} of the planet transit, especially for large impact parameters
 when a small change in inclination significantly affects the length of the 
 transit chord.

Due to its four year mission lifetime, \Kepler\ is not optimal for observing 
 these secular effects, which occur on timescales of order
\begin{equation}
\tau \sim \frac{M_\star}{M_p}P_{tr}
\end{equation}
with $M_\star$ the mass of the host star, $M_p$ the perturber's mass, and
 $P_{tr}$ the period of the transiting planet. 
For a planet in a 10-day orbit perturbed by a $2 M_J$ object, this cycle is 
 of order 10 years. 
 However, a repurposed \Kepler\ working in concert with
 TESS or a future TESS-like mission would be the ideal instrument for this 
 study. 
\Kepler\ will be able to measure the transit durations of large, warm 
 transiting planets in a few fields in 2014. 
Moreover, with an increased cadence the precision of observations will be 
 enhanced over the work currently accomplished in the \Kepler\ field. 
Observing transits each year will enable a search for transit duration 
 variations (TDVs). 
When these fields are revisited in 2018-2019 with TESS, we will immediately have
 at least a five year baseline to compare transit durations against. 
\textbf{This is a science objective that could not be carried out by TESS or 
 \Kepler\ alone}. 
A search for TDVs will enable key outstanding questions about the architecture
 of exoplanetary systems to be answered;  there are not any missions current or
 planned that will be able to study the flatness of exoplanetary systems as 
 well as the combination of a repurposed \Kepler\ and TESS.

Additionally, a shorter cadence which is more sensitive to transit duration 
 variations might aid in the Hunt for Exomoons with \Kepler\
 \citep{Kipping12}, which requires sensitive transit timing and duration 
 observations.

\subsection{Masses of Transiting Planets}
An unexpected success of the \Kepler\ mission has been the discovery of many 
 systems with tightly-packed inner planets 
 \citep[hereafter STIPs, e.g.][]{Boley13}. 
Ten percent or more of stars appear to have planets a few Earth radii in size
 with periods smaller than 20 days. 
These systems appear to form around stars of a variety of spectral types, from 
 Solar-type stars \citep[Kepler-11,][]{Lissauer11} to mid M dwarfs 
 \citep[Kepler-42,][]{KOI961}.
While their existence is unquestioned, their formation is uncertain. 
\cite{Boley13} propose \textit{in-situ} formation via aerodynamic drift, and 
 \cite{Chiang13} also suggest \textit{in-situ} formation is reasonable. 
Meanwhile, \cite{Swift13} and \cite{Cossou13} propose migration of 
 planetary embryos to form such systems.
To better understand the formation and evolution of these systems, it 
 is imperative to understand their mass (and thus density) distributions. 
The most effective method to determine masses of small transiting planets 
 to date has been the characterization of transit timing variations (TTVs) 
 \citep[e.g.][]{Fabrycky12}, the effects of gravitational perturbations
 between planets near each conjunction. 

Far and away the most common type of TTV signals observed with 
 \Kepler\ are variations that occur on the timescale of planetary conjunctions
 for near-resonant systems (the ones described above). 
A mission that observes systems for approximately one month per year is 
 suboptimal for detailed characterization of this type of TTV for specific 
 systems. 
However, all hope is not lost.
In the most commonly analyzed case where both near-resonant planets transit,
 the period of the TTV signal is known and equal to the period of conjunctions,
 leaving the only free parameters the amplitude of the TTV signal and its phase,
 assuming free eccentricity in the system is negligible \citep{Wu13}.
As the typical period of these conjunctions is 1-2 years, one quarter of
 observations is not enough to determine the masses of the transiting planets. 
However, return observations of these systems over multiple years with \Kepler\
 and eventually with TESS can enable a full cycle to be observed, allowing for 
 masses to be estimated. Even when a full cycle can not be observed, strong 
 upper limits can be placed on the masses of the planets from TTV nondetections.
Additionally, an increased sensitivity to transit timing variations would 
 provide for an increased ability to confirm planetary systems without 
 radial velocity follow-up, as the presence of dynamical effects can 
 immediately cause false positive scenarios to be discarded.

\Kepler\ was, fundamentally, a statistical mission. 
A study such as this will help us better understand the statistics of TTV 
 systems. 
While we will likely not be able to characterize systems as well as can be
 presently accomplished \citep[e.g. KOI-142,][]{Nesvorny13}, statistical
 analyses that would benefit from an increased number of dynamically 
 interacting systems can be undertaken.
If the updated \Kepler\ mission collects four pointings each of 30,000 stars, 
 then of the 120,000 stars observed, 12,000 would be expected to host STIPs
 and approximately 600 of these systems would be expected to transit, allowing 
 for the discovery of hundreds of near-resonant systems.
Adding numbers such as these to the current sample will allow a significant 
 increase in our understanding of the properties of these systems and provide
 insight into their formation.

\section{Other Benefits}
\label{AS}
In some ways, the proposed mission is similar to a mini-TESS. 
There are considerable benefits to such a strategy. 
Many have been outlined above; the potential to open up TESS to searching for 
 TDVs and TTVs over 400 square degrees of the sky cannot be overstated. 
TESS will also have the ability to search for longer-period planets in these 
 fields when its data is combined with \Kepler\ data.
It is worth mentioning that this proposed survey will only cover one percent 
 of the sky, and will thus not impinge on TESS' discovery space in any 
 significant way. 
It is of our opinion that the benefits to observing these fields in the 
 first half of this decade as a precursor to TESS only serves to enhance 
 the future mission, as opposed to detracting from it.

Observing brighter stars at a higher cadence will allow asteroseismic 
 studies of more stars than were observable with the completed
  \Kepler\ mission. 
Long cadence data can be used to characterize global properties of giant stars
 that display asteroseismic oscillations. 
The Nyquist frequency for a cadence of 29.4 minutes is 283uHz,
 which corresponds to stars at the base of the red giant branch on the
 HR diagram. 
Increasing the cadence would impact the population of stars available for
 asteroseismic analysis, bringing bright sub-giants into the asteroseismic
 regime (Chaplin, W. J., private communication). 
Since the magnitude of stellar pulsations decreases with increasing
 stellar density, brighter sub-giant targets would need to be selected
 in order to reach the current measurement precision for giant stars.

An additional benefit to focusing on brighter stars is the increased ability 
 for radial velocity (RV) followup. 
If our proposed observing strategy is undertaken, there will
 be four new \Kepler\ fields, two near a declination of +20 degrees and two 
 near -20 degrees. 
These are ideally placed for follow up radial velocity work
 by existing telescopes in Hawaii and Chile, as well as future 30-meter class
 telescopes. 
The faintest stars in the \Kepler\ field are too faint to be observed 
 efficiently on existing telescopes.
With observing time as competitive as it will likely be on the proposed
 giant telescopes, observations \textit{en masse} of the faintest transiting 
 exoplanets will likely not be achievable with these facilities.
Therefore, a focus on brighter stars (perhaps $K_p < 15$ for solar-type stars) 
 will enable more efficient and successful RV follow up.
 
Finally, and perhaps most significantly, such a mission will allow potentially 
 interesting objects to be found before the launch of the James Webb Space 
 Telescope (JWST). 
Scheduled to launch in 2018, JWST will be placed at 
 the inaccessible L2 Lagrange point, making it necessarily a fixed-length 
 mission. 
Moreover, by this time the \textit{Spitzer} telescope will have drifted too 
 far from Earth to provide useful data.
TESS is scheduled to launch at approximately the same time as JWST, but 
 space telescopes do occasionally run into delays for various reasons\footnote{see also: JWST.}.
If TESS were to be delayed, and JWST ran for only its nominal five year 
 mission, it is conceivable that many of TESS' most interesting discoveries 
 will occur at a time when there are no available infrared space facilities 
 available for follow up work.
Certainly, even if both telescopes proceed according to plan, time will be 
 limited between publication of TESS' results and the end of the JWST mission.
This proposal, will find interesting transiting systems across 
 the sky well before the launch of JWST, allowing 
 considerable time to plan observations with the future observatory before 
 launch and assuring that JWST will have an abundance of planets to 
 characterize.

\section{Summary and Conclusions}
\label{SC}
Our strategy outlined here is one possible observing strategy. 
Of course, there are considerable benefits to continuing to observe the 
 current \Kepler\ field. We assume these benefits will be discussed in other 
 white papers. 
Unfortunately, there is no guarantee our methods (described fully in Hogg 
 et al.) will be able to be applied to the \Kepler\ field because of the large
  torque induced on the telescope along the Y axis by solar radiation pressure.
In the case where new fields must be chosen, we recommend selecting four 
 fields subject to the constraints of \textsection\ref{Targets}. 
This strategy, in addition to uncovering 800 or more new planets, 
 will provide the best chance to improve constraints on the frequency of 
 habitable zone planets as a function of stellar spectral type, one of the 
 key goals of the \Kepler\ mission. 
It will also enable new transit timing and duration studies, allowing us to 
 probe the flatness of exoplanetary systems with better precision.
This strategy will also turn TESS into a transit timing machine in these fields,
 creating a 5 year baseline for dynamical studies where otherwise none would 
 exist.
Moreover, the detection of 800 new planets well before the launch of JWST is 
 significant as it allows time to plan follow-up analyses of transiting 
 exoplanets with this telescope across the sky well before its launch.
If our image modeling techniques are successful, we feel the strategy outlined
 herein provides a unique opportunity for the contributions of \Kepler\ 
 to continue unhindered for the next decade through a combination of its own 
 observations and those planned in the future by TESS. 

\section{Acknowledments}
Much like its namesake, \Kepler\ has been instrumental\footnote{Pun not 
 intended.} in ushering in an astronomical revolution.
It is our great pleasure to have used data from the \Kepler\ telescope since 
 its launch; we acknowledge the hard work put in by this team both in building 
 the telescope and providing data and support since 2009.
BTM is supported by the NSF Graduate Research Fellowship grant DGE-1144469.
DWH, RF, and DFM are all partially supported by NSF grant IIS-1124794.
MH acknowledges support from the European Research Council in the form of a 
 Starting Grant with number 240672.

\end{document}